\documentclass[article,twocolumn,amsmath,amssymb,superscriptaddress,floatfix,nofootinbib,10pt]{revtex4}
\usepackage{times}
\usepackage{amssymb,amsbsy,amsmath,amsfonts}
\usepackage{graphicx}
\usepackage{float}
\usepackage{color}
\usepackage{morefloats}
\usepackage{rotating}
\usepackage{srcltx}
\usepackage{slashed}
\usepackage{multirow}
\usepackage{verbatim}
\usepackage{hyperref}
\usepackage{tabularx}
\usepackage{bm}
\allowdisplaybreaks[4]

\usepackage{bbding}
\usepackage{threeparttable}

\newcommand{\PreserveBackslash}[1]{\let\temp=\\#1\let\\=\temp}
\newcolumntype{C}[1]{>{\PreserveBackslash\centering}p{#1}}
\newcolumntype{R}[1]{>{\PreserveBackslash\raggedleft}p{#1}}
\newcolumntype{L}[1]{>{\PreserveBackslash\raggedright}p{#1}}

\begin{document}

\title{How much is the compositeness of a bound state constrained by $a$ and $r_0$? The role of the interaction range}

\author{Jing Song}
\email[E-mail me at: ]{Song-Jing@buaa.edu.cn}
\affiliation{School of Physics, Beihang University, Beijing, 102206, China}
\affiliation{Departamento de Física Teórica and IFIC, Centro Mixto Universidad de Valencia-CSIC Institutos de Investigación de Paterna,
Aptdo.22085, 46071 Valencia, Spain}

\author{L.R.Dai}
\email[E-mail me at: ]{ dailianrong@zjhu.edu.cn}
\affiliation{Departamento de Física Teórica and IFIC, Centro Mixto Universidad de Valencia-CSIC Institutos de Investigación de Paterna,
Aptdo.22085, 46071 Valencia, Spain}
\affiliation{School of science, Huzhou University, Huzhou, 313000, Zhejiang, China}

\author{ E.Oset}
\email[E-mail me at: ]{oset@ific.uv.es}
\affiliation{Departamento de Física Teórica and IFIC, Centro Mixto Universidad de Valencia-CSIC Institutos de Investigación de Paterna,
Aptdo.22085, 46071 Valencia, Spain}

\begin{abstract}
We present an approach that allows one to obtain information on the compositeness of molecular states from combined information of the scattering length of the hadronic components, the effective range, and the binding energy. We consider explicitly the range of the interaction in the formalism and show it to be extremely important to improve on the formula of Weinberg obtained in the limit of very small binding and zero range interaction. The method allows obtaining good information also in cases where the binding is not small. We explicitly apply it to the case of the deuteron and the $D^{*}_{s0}(2317)$ and $D^{*}_{s1}(2460)$  states and determine simultaneously the value of the compositeness within a certain range, as well as get qualitative information on the range of the interaction.

\end{abstract}


\date{\today}

\maketitle
\section{Introduction}
The determination of the compositeness, or molecular component of physical hadronic states has been the subject of multiple discussions starting from the pioneer work of Weinberg~\cite{PhysRev.137.B672}. In one of the most popular variants of the relationship of low energy scattering observables to the compositeness $X_{W}$, or probability to have a molecular state of two hadrons, one finds  
\begin{align}\label{1_1}
    a=R\bigg[\frac{2X_W}{1+X_W}+O(\frac{R_{\mathrm{typ}}}{R})\bigg]
\end{align}
\begin{align}\label{1_2}
    r_0=R\bigg[-\frac{1-X_W}{X_W}+O(\frac{R_{\mathrm{typ}}}{R})\bigg]
\end{align}
$R=1/\sqrt{2\mu B}$, $\mu=\frac{m_1m_2}{m1+m_2}$, $B=E_{\mathrm{th}}-E_0$, $E_0$ and $E_{\mathrm{th}}$ the energy of the bound state and the threshold energy of the two particles with mass $m_1$and $m_2$, with $a$, $r_0$ the scattering length and effective range in the approximation of the scattering matrix,
\begin{align}\label{1_3}
    f=\frac{1}{k\cot\delta-ik}\approx \frac{1}{-\frac{1}{a}+\frac{1}{2}r_0k^2-ik}
\end{align}

Eqs.~(\ref{1_1}),~(\ref{1_2}) hold in the limit of small binding energy $B$ and $O(\frac{R_{\mathrm{typ}}}{R})$ are corrections to the formula stemming from a typical scale $R_{\mathrm{typ}}$ related to the range of the interaction.

If we invert Eqs.~(\ref{1_1}),~(\ref{1_2}), we find 
\begin{align}\label{2_1}
    X_W=\frac{a}{2R-a}+O(\frac{R_{\mathrm{typ}}}{R})
\end{align}
\begin{align}\label{2_2}
    X_W=\frac{R}{R-r_0}+O(\frac{R_{\mathrm{typ}}}{R})
\end{align}
and one also defines $Z=1-X_W$ as the amount of non molecular component of the bound state.

If we apply these rules to the deuteron, where for $I=0$, $J=1$, we have
\begin{align}\label{2_3}
    &a=5.419(7)\mathrm{fm},\\\nonumber
    &r_0=1.766(8)\mathrm{fm},\\\nonumber
    &B=2.224575(9)\mathrm{MeV}
\end{align}
one finds the well known surprise~\cite{Guo:2017jvc,Esposito:2021vhu,Li:2021cue} that 
\begin{align}\label{2_4}
    &X_W=1.68\quad \mathrm{from\quad Eq.~(\ref{2_1})},\\\nonumber
    &X_W=1.69 \quad \mathrm{from\quad Eq.~(\ref{2_2})}
\end{align}
with an unacceptable result since the actual compositeness, $X$, should be $X\leq1$. Eq.~(\ref{1_2}) also yields an unacceptable result, which forces $r_0$ to be negative if $X_W$ has an acceptable value around $1$, Yet, the test is accepted as a good guess for a molecular $np$ structure of the deuteron because the alternative is even worse: For $X=0.1$, Eq.~(\ref{1_2}) yields $r_0=-38.86\mathrm{fm}$, very far from the relative small number of Eq.~(\ref{2_3}). Indeed, this is the argument of Weinberg quoting ``The true token that the deuteron is composite is that $r_0$ is small and positive rather than large and negative''. Since for different reasons the molecular picture of the deuteron is acceptable, one can only conclude that the corrections $O(\frac{R_{\mathrm{typ}}}{R})$ in Eqs.~(\ref{1_1}),~(\ref{1_2}) are very large. However, they are difficult to quantify in terms of low energy variables.

Lots of energies have been devoted to improve the Weinberg conditions, Eqs.~(\ref{1_1}),~(\ref{1_2}), and understand the meaning of the compositeness in the presence of coupled channels or energy dependent potentials~\cite{BARU200453,Gamermann:2009uq,Hyodo:2011qc,Baru:2010ww,Hanhart:2011jz,Aceti:2012dd,Hyodo:2013iga,Sekihara:2014kya,Hanhart:2014ssa,Guo:2015daa,Sekihara:2015gvw,Kamiya:2015aea,Sekihara:2016xnq,Kamiya:2016oao,Matuschek:2020gqe,Guo:2017jvc}. Very recently the role of the range of the interaction has been revisited in~\cite{Kinugawa:2021ykv,Li:2021cue}. In~\cite{Kinugawa:2021ykv} a simple field theoretical approach is done containing a point like interaction and a derivative coupling term. A different version is also studied incorporating an additional field that couples to the two body scattering state. In these cases it is shown that the effective range $r_0$ originates from the derivative coupling interaction or from the channel coupling to the bare state. It is suggested that $R_{\mathrm{typ}}$ in Eqs.~(\ref{1_1}),~(\ref{1_2}) should be $R_{\mathrm{typ}}=\mathrm{max}[R_{\mathrm{int}},R_{\mathrm{eff}}]$, where $R_{\mathrm{int}}$ is $1/\Lambda$($\Lambda$, cutoff or range of the interaction) and $R_{\mathrm{eff}}$ is a length scale in the effective range expansion of Eq.~(~\ref{1_3}). The developments are formal and some examples are worked out in~\cite{Kinugawa:2021ybb}.

A different approach is followed in~\cite{Li:2021cue}. The Low equation~\cite{PhysRev.137.B672} for the scattering matrix is used and the non pole term is neglected. Some form factors are introduced and an expression for $X$ is obtained which depends on phase shifts obtained with some approximations, which has the virtue of being smaller than $1$. A discussion is done on how the results can depend on the form factors assumed, which neglecting the non pole term and making a separable ansatz for the $T$-matrix can be worked out up to order $O(p^2)$. In Ref.~\cite{Baru:2021ldu} the effect of coupled channels and the range of the interaction are also addressed and shown to be relevant to go beyond Eqs.~(\ref{1_1}), (\ref{1_2}).

The purpose of the present work is different. We share with the former works the idea that the range of the interaction is relevant in the determination of $X$. The aim, however, is to see how much information can we obtain from the combined knowledge of $a$, $r_0$ and the binding, not knowing exactly which is the range of the interaction. For this purpose we start with a formalism that takes into account the range of the interaction and incorporates a possible energy dependence of the potential. With the information on the binding, $a$ and $r_0$ we investigate the range of values of $X$ that the formalism provides. As we shall see, we can obtain a more accurate prediction for $X$ for the deuteron, and we can obtain a positive $r_0$, while at the same time we obtain a qualitative information on the range of the interaction. We also discuss two more states, the $D^{*}_{s0}(2317)$ and $D^{*}_{s1}(2460)$, which were studied in detail from a lattice QCD perspective in~\cite{MartinezTorres:2014kpc}, and the value of $X$, $a$, $r_0$ were determined. The lattice data of~\cite{MartinezTorres:2014kpc} contain more information than just $a$, $r_0$ and the binding, which allowed one to determine $X$ with relative precision. In the present work we shall exploit how much information we can get on $X$ from the knowledge of $a$, $r_0$ and $B$ alone.

\section{formalism}
\subsection{Scattering matrix with a separable potential}
The formalism will be based on a derivation of the scattering matrix using a separable potential. We state that at the beginning such that the approximations done and the limitations are clear. We shall discuss later on how accurate this assumption can be. We follow closely the work of Ref.~\cite{Gamermann:2009uq} for the derivation, and do it for one channel for simplicity. The extension to coupled channels is trivial and is also done in Ref~\cite{Gamermann:2009uq}. We start from a potential written in momentum space as
\begin{align}\label{1_1_1}
   \langle\textbf{p}'|V|\textbf{p}\rangle= V(\textbf{p}',\textbf{p})=V\theta({q_{\mathrm{max}}-p'})\theta({q_{\mathrm{max}}-p})
\end{align}
where $p'$, $p$ are $|\textbf{p}'|$, $|\textbf{p}|$ respectively. It is clear from the beginning what is the meaning of $q_{\mathrm{max}}$. It gives the range of the potential in momentum space. Its inverse would provide the range of the interaction in coordinate space. We have chosen a sharp cut off in the interaction, but the results are easily extended to any other type of separable potential. Next we solve the Bethe Salpeter equation with this potential to obtain the $T$-matrix.

\begin{align}
  &\langle \textbf{p}'|T|\textbf{p}\rangle =T(\textbf{p}',\textbf{p})\\\nonumber
   =&V(\textbf{p}',\textbf{p})
   +i\int\frac{d^4q}{(2\pi)^4}\frac{V(\textbf{p}',\textbf{q})}{q^2-m_1^2+i\epsilon}\frac{T(\textbf{q},\textbf{p})}{(P-q)^2-m_2^2+i\epsilon}
\end{align}
where $m_1$, $m_2$ are the masses of the interacting particles (we use the meson formalism) and $P$ is  the total four momentum of the pair. The $q^0$ integration is readily done using Cauchy's residues with the result
\begin{align}\label{2_1_1}
  & T(\textbf{p}',\textbf{p})\\\nonumber
   =&V(\textbf{p}',\textbf{p})
   +\int\frac{d^3\textbf{q}}{(2\pi)^3}V(\textbf{p}',\textbf{q})T(\textbf{q},\textbf{p})\frac{w_1(\textbf{q})+w_2(\textbf{q})}{2w_1(\textbf{q})w_2(\textbf{q})}\\\nonumber
   &~~~~~~~~~~~~~~~~~~~~~~\times\frac{1}{(P^0)^{2}-(w_1(\textbf{q})+w_2(\textbf{q}))^2+i\epsilon}
\end{align}
with $w_i(\textbf{q})=\sqrt{\textbf{q}^2+m^2_i}$, $(P^0)^{2}=s$.

By expanding Eq.~(\ref{2_1_1}) in a power series we see that in all terms we have
\begin{align}
   \theta({q_{\mathrm{max}}-p'}|)\theta({q_{\mathrm{max}}-q})\theta({q_{\mathrm{max}}-p})
\end{align}
with $q=|\textbf{q}|$, and hence we factorize $\theta({q_{\mathrm{max}}-p'})\theta({q_{\mathrm{max}}-p})$ outside the integrand with the result that\\
\begin{align*}
   T(\textbf{p}',\textbf{p})=\theta({q_{\mathrm{max}}-p'})\theta({q_{\mathrm{max}}-p})T
\end{align*}
where 
\begin{align}\label{2_2_1}
T=V+VGT
\end{align}
with
\begin{align}\label{2_3_1}
   G(s)=\int_{|\textbf{q}|< q_{\mathrm{max}}}&\frac{d^3\textbf{q}}{(2\pi)^3}\frac{w_1(\textbf{q})+w_2(\textbf{q})}{2w_1(\textbf{q})w_2(\textbf{q})}\\\nonumber
   \times&\frac{1}{s-(w_1(\textbf{q})+w_2(\textbf{q}))^2+i\epsilon}
\end{align}
Eq.~(\ref{2_2_1}) becomes then an algebraic equation where
\begin{align}\label{5_2}
    T=[1-VG]^{-1}V
\end{align}

This is the equation used for instance in studies of the chiral unitary approach~\cite{Oller:1997ti}  for meson meson interaction, from where poles and couplings of bound states are obtained. 

The results obtained here are usually presented from a different perspective in~\cite{Oller:1997ti,Oset:1997it,Oller:2000fj}. One can reach the same conclusion about the $T$-matrix by assuming that one can factorize on shell the $V$ and $T$ matrices inside the integral of Eq.~(\ref{2_1_1}), obtaining Eq.~(\ref{2_2_1}), and the G-function is then regularized with a cut off. The on shell factorization is justified in~\cite{Oller:1997ti,Oset:1997it} showing that the contribution of the off shell part of the potential in Eq.~(\ref{2_1_1}) using chiral Lagrangians can be reabsorbed in the on shell potential itself. A different justification can be found using a dispersion relation as done in~\cite{Oller:2000fj}. If one neglects the energy dependence of the contribution from the left hand cut, the on shell factorization also arises. In Ref.~\cite{Oller:1998zr} it is shown that this is a very good approximation for meson meson interactions in the range of energies where the low energy resonance appear. Further discussion on this issue is shown in the recent review~\cite{Meng:2022ozq}. The success of the chiral unitary approach to obtain the spectrum of low lying resonances and describe features of many reactions has been reported in some reviews~\cite{Oller:2000ma,Oset:2016lyh}.

We have shown the equivalence of using a separable potential of Eq.~(\ref{1_1_1}) and the on shell factorization used in the chiral unitary approach. For the present work the perspective of the separable potential is better because it allows to identify $q_{\mathrm{max}}$ with the range of the interaction from the very beginning. 

Eq.~(\ref{5_2}) is generalized to coupled channels with exactly the same form expect the 1 is the identity matrix in the dimension of the number of channels, $n$. $V_{ij}$ is the $n\times n$ transition potential  matrix and $G$ is the diagonal $G$-matrix with the $G$-function for each of the channels. In the next subsection we shall work with two channels assuming that the two channels account for the whole wave function of a certain state, and then will eliminate one of the channels, investigating how we can obtain the probability of the remaining channel working with that channel alone.

\subsection{Formalism for the meson meson interaction in two channels}
Let us start with a meson meson interaction and the formalism employed in the studies of the chiral unitary approach. To put the problem in perspective let us start with a two channel problem, in which one channel is more important than the other and can lead to a bound state with this interaction. Let the interaction be given by the matrix 
\begin{align}\label{5_1}
    V=
    \left(
  \begin{array}{cc}
    V_{11} & V_{12} \\ 
    V_{12} & 0 \\
  \end{array}
\right),
\end{align}
where for simplicity we have made $V_{22}=0$. The formalism can be generalized to more coupled channels keeping all $V_{ij}$ terms~\cite{Hyodo:2013nka}. We assume $V_{ij}$ in Eq.~(\ref{5_1}) to be energy independent. The $T$ matrix is then given by Eq.~(\ref{5_2}), where $G=\mathrm{diag}(G_l)$ is the loop function of two mesons, with the cut off method as shown in Eq.~(\ref{2_3_1}), where $G_l$ is the $G$-function evaluated for each of the channels.

\begin{figure}[H]
  \centering
  \includegraphics[width=0.45\textwidth]{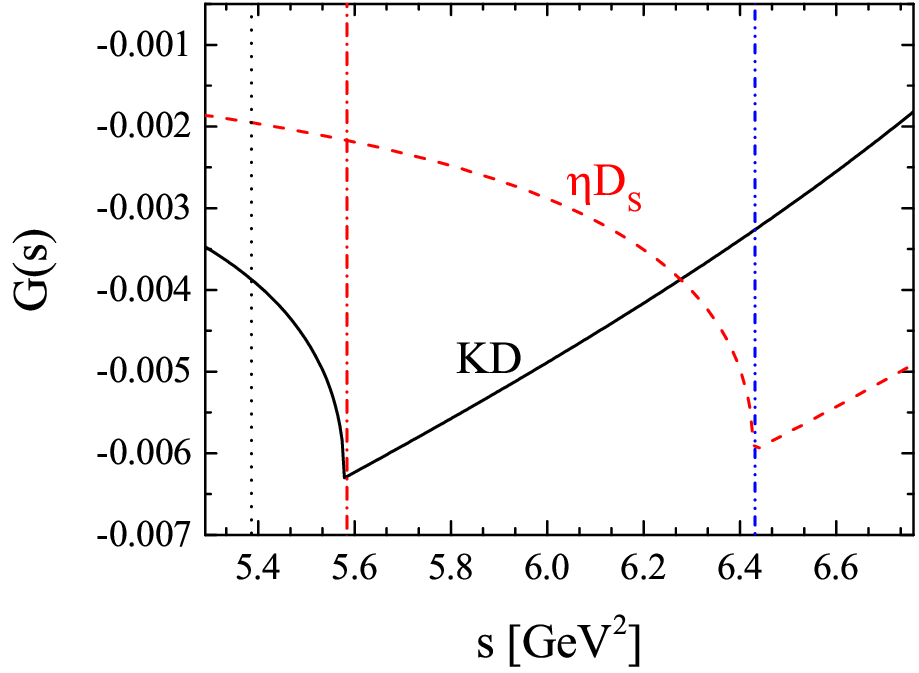}
  \caption{The $G(s)$ function for the $DK$ and $\eta D_s$ channels at $q_{\mathrm{max}}=650$~MeV. The black solid line and red dashed line represent the $DK$ and $\eta D_s$ channels respectively, with the black dotted line, red dashed dotted line and blue dashed-dotted-dotted line corresponding to $\sqrt{s}=2317$~MeV, the threshold of the $KD$ channel, and the threshold of the  $\eta D_s$ channel.}\label{G_s}
\end{figure}

Let us assume that Eq.~(\ref{5_2}) has a pole at $s_R$ corresponding to a bound state in channels $1$ and $2$. We define the couplings of the state to the channels $1$ and $2$ as
\begin{align*}
  g_1^2=\lim\limits_{s-s_R}(s-s_R)T_{11},\qquad g_2^2=\lim\limits_{s-s_R}(s-s_R)T_{22}
\end{align*}
and we can explicitly prove that (see section 5 of ~\cite{Aceti:2014ala})
\begin{align}\label{6_1}
    -g_1^2\frac{\partial G}{\partial s}|_{s=s_R}-g_2^2\frac{\partial G}{\partial s}|_{s=s_R}=P_1+P_2=1
\end{align}
where $P_1$, $P_2$ represent the probability to find the state in the channel $1$ and $2$. A general proof for many channels can be seen in ~\cite{Gamermann:2009uq,Hyodo:2013nka}. As shown in section 6 of Ref.~\cite{Aceti:2014ala}, one can construct an effective potential for channel $1$
\begin{align}\label{6_2}
    V_{\mathrm{eff}}=V_{11}+V_{12}^2G_2
\end{align}
such that $T_{11}$ and $g_1^2$ are the same in channel $1$ as a single channel using $V_{\mathrm{eff}}$  than with the two channels problem, and hence $-g_1^2\frac{\partial G_1}{\partial s}$ is still the probability to the find channel $1$ in the molecular state. Note that the price we payed to eliminate channel $2$ is that the effective potential in channel $1$, Eq.~(\ref{6_2}) is now energy dependent because $G_2$ depends on the energy. Now we have with one channel 
\begin{align}
    T_{11}=[1-V_{\mathrm{eff}}G_1]^{-1}V_{\mathrm{eff}}=\frac{1}{V_{\mathrm{eff}}^{-1}-G_1}
\end{align}
\begin{align*}
  g_1^2=\lim\limits_{s-s_R}(s-s_R)T_{11}=\frac{1}{\frac{\partial{V_{\mathrm{eff}}^{-1}}}{\partial s}|_{s_R}-\frac{\partial G}{\partial s}|_{s_R}}  
\end{align*}
and hence
\begin{align}\label{6_3}
    -g_1^2\frac{\partial G_1}{\partial s}|_{s_R}&=P_1,\\\nonumber
    g_1^2\frac{\partial{V_{\mathrm{eff}}^{-1}}}{\partial s}|_{s_R}&=-g_1^2\frac{1}{V_{\mathrm{eff}}^{2}}\frac{\partial{V_{\mathrm{eff}}}}{\partial s}=1-P_1\equiv P_2\equiv Z
\end{align}
To have a feeling of the energy dependence of $V_{\mathrm{eff}}$, we particularize to a case that will be studied later, the $D^{*}_{s0}(2317)$, which qualifies as a molecular state of $DK$ and $\eta D_s$, mostly $DK$~\cite{Kolomeitsev:2003ac,Guo:2006fu,Gamermann:2006nm,Guo:2009ct,Wang:2012bu,Liu:2012zya,Altenbuchinger:2013gaa}. In Fig.~\ref{G_s} we show the $G(s)$ function for the $DK$ and $\eta D_s$ channels. We can see that around the energy of the $D^{*}_{s0}(2317)$, between $\sqrt{s}=2317$~MeV and the $DK$ threshold, the $G_{\eta D_s}(s)$ function is very well represented by a linear function in $s$. 
We shall then assume as a general rule that the $G$ function of possible missing channels, when studying a given one, can be approximated by a linear function of $s$ in the neighborhood of the pole of the studied channel, such that we can well approximate $V_{\mathrm{eff}}$ by
\begin{align}\label{7_1}
   V_{\mathrm{eff}}=V_{0}+\beta(s-s_{0})
\end{align}
with $\beta$ negative, as seen in Fig.~\ref{G_s}. If we eliminated some channel with a threshold below the threshold of the studied channel, $\beta$ would be positive. It is possible to tackle also this case, but in this case it would be convenient to consider explicitly the possible decay channels. Hence, we stick to Eq.~(\ref{7_1}) with $\beta\leq 0$.

The scattering matrix with potential $V_{\mathrm{eff}}$ of Eq.~(\ref{7_1}) is given for the one channel system that we wish to investigate by
\begin{align}\label{8_1}
    T(s)=\frac{1}{[V_{0}+\beta(s-s_{0})]^{-1} -G(s)}
\end{align}
and we impose that it has a pole at $s_0$, giving rise to a state whose nature we want to investigate. We will have 
\begin{align}\label{8_11}
    V_{0}^{-1} -G(s_0)=0,\qquad V_{0}=\frac{1}{G(s_0)}
\end{align}
Thus,
\begin{align}\label{8_2}                          T(s)    =&\frac{1}{[\frac{1}{G(s_0)}+\beta(s-s_0)]^{-1} -G(s)}
\end{align}
where we have eliminated the unknown $V_0$.

It is easy to establish the connection of this expression with the current amplitude of Quantum Mechanics by restoring the normalization of $T(s)$, governed by the normalization of the $G(s)$ function of Eq.~(\ref{2_3_1}). We have 
\begin{align}\label{8_3}
    \frac{1}{C\bigg\{[\frac{1}{G(s_0)}+\beta(s-s_0)]^{-1} -G(s)\bigg\}}\approx\frac{1}{-\frac{1}{a}+\frac{1}{2}r_0k^2-ik}
\end{align}
with $C$ a normalization constant. In the denominator of the right hand side there would be more terms in an exact solution, but we are only interested in obtaining $a$ and $r_0$ from our $T$ matrix.
Since 
\begin{align}
   \mathrm{Im}G=&
   -\frac{1}{8\pi}\frac{k}{\sqrt{s}}
\end{align}
we can establish the connection
\begin{align}
    &8\pi\sqrt{s}\bigg\{[\frac{1}{G(s_0)}+\beta(s-s_0)]^{-1} -\mathrm{Re}G(s)\bigg\}+ik\\\nonumber
        \approx &\frac{1}{a}-\frac{1}{2}r_0k^2+ik
\end{align}
or
\begin{align}\label{8_4}
    &8\pi\sqrt{s}\bigg\{[\frac{1}{G(s_0)}+\beta(s-s_0)]^{-1} -\mathrm{Re}G(s)\bigg\}\\\nonumber
        \approx &\frac{1}{a}-\frac{1}{2}r_0k^2
\end{align}

From Eq.~(\ref{8_4}), we obtain two equations evaluating the expression on the left at threshold and its derivative with respect to $k^2$ at threshold for positive increases of $s$. We have 
\begin{align}\label{9_1}
    8\pi\sqrt{s_\mathrm{th}}\bigg\{[\frac{1}{G(s_0)}+\beta(s_{\mathrm{th}}-s_0)]^{-1} -\mathrm{Re} G(s_{\mathrm{th}})\bigg\}
        =&\frac{1}{a}
\end{align}
and taking into account that
  $s=(w_1(k)+w_2(k))^2$, $w_1(k)=\sqrt{m_1^2+k^2}$, $w_2(k)=\sqrt{m_2^2+k^2}$,
\begin{align}
    \frac{\partial }{\partial k^2}&=\frac{\partial }{\partial s}\frac{\partial s}{\partial k^2}\\\nonumber
    &=2(w_1(k)+w_2(k))(\frac{1}{2w_1(k)}+\frac{1}{2w_2(k)})\frac{\partial }{\partial s}\\\nonumber
    &=\frac{(w_1(k)+w_2(k))^2}{w_1(k)w_2(k)}\frac{\partial }{\partial s}.
\end{align}
we obtain
\begin{align}\label{9_2}
    \frac{1}{2\sqrt{s_\mathrm{th}}}8\pi\bigg[&[\frac{1}{G(s_0)}+\beta(s_\mathrm{th}-s_0)]^{-1}\\\nonumber
    & -\mathrm{Re}G(s)_\mathrm{th}\bigg]\frac{s}{w_1(k)w_2(k)}|_{s_{\mathrm{th}}}\\\nonumber
    +
    8\pi\sqrt{s_\mathrm{th}}\bigg[&-\beta[\frac{1}{G(s_0)}+\beta(s_\mathrm{th}-s_0)]^{-2}\\\nonumber
    &-\frac{\partial{\mathrm{Re}[G(s)]}}{\partial s}|_{s^{+}_{\mathrm{th}}}\bigg]\frac{s}{w_1(k)w_2(k)}|_{s_{\mathrm{th}}}
    =    
    -\frac{1}{2}r_0
\end{align}
From Eq.~(\ref{9_1}) we get the value of $\beta$ from $a$, as
\textbf{\begin{align}\label{9_3}
    \beta=&\frac{1}{s_{\mathrm{th}}-s_0}\bigg\{[\frac{1}{a}\frac{1}{8\pi}\frac{1}{\sqrt{s_{\mathrm{th}}}}+\mathrm{Re}G(s_{\mathrm{th}})]^{-1}-\frac{1}{G(s_0)}\bigg\}
\end{align}}
Note that using Eqs.~(\ref{6_3}),~(\ref{7_1}),~(\ref{8_11}) we have 
\begin{align}\label{9_4}
    P_2=1-P_1=Z=-g^2G(s_0)^2\beta
\end{align}
with 
\begin{align}\label{9_5}
        g^2&=\lim\limits_{s-s_0}(s-s_0)T(s)\\\nonumber
        &=\lim\limits_{s-s_0}\frac{s-s_0}{[\frac{1}{G(s_0)}+\beta(s-s_0)]^{-1} -G(s)}\\\nonumber
        &=\frac{1}{-(\frac{1}{G(s_0)})^{-2}\beta-\frac{\partial G}{\partial s}|_{s_0}}=\frac{1}{-G(s_0)^2\beta-\frac{\partial G}{\partial s}|_{s_0}}
\end{align}
where we have applied L'Hospital's rule to calculate the limit. Eq.~(\ref{9_4}) with $\beta \leq 0$ will always guarantee that $Z$ is positive, as it should be (note that $\frac{\partial G}{\partial s}|_{s_0}< 0$).

Eq.~(\ref{9_3}), with the knowledge of $a$ and $q_{\mathrm{max}}$ and the position of the pole will give the value of $\beta$  and by means of Eqs.~(\ref{9_4}),~(\ref{9_5}), we will get the value of $Z$, the complement of the compositeness of the state that we study. With the value of $\beta$ obtained we can go to Eq.~(\ref{9_2}) and establish that 
\begin{align}\label{10_1}
   R_0\equiv -\frac{1}{2\sqrt{s_\mathrm{th}}}16\pi\bigg[&[\frac{1}{G(s_0)}+\beta(s_\mathrm{th}-s_0)]^{-1}\\\nonumber &-\mathrm{Re}G(s_\mathrm{th})\bigg]\frac{s}{w_1(k)w_2(k)}|_{s_{\mathrm{th}}}\\\nonumber
    +16\pi\sqrt{s_\mathrm{th}}\bigg[&\beta[\frac{1}{G(s_0)}+\beta(s_\mathrm{th}-s_0)]^{-2}\\\nonumber
    &+\frac{\partial{\mathrm{Re}[G(s)]}}{\partial s}|_{s^{+}_{\mathrm{th}}}\bigg]\frac{s}{w_1(k)w_2(k)}|_{s_{\mathrm{th}}}
    =    
    r_0
\end{align}
where $R_0$ is the theoretical value that our approach provides for $r_0$.

Given $a$ and $s_0$, $R_0$ in Eq.~(\ref{10_1}) is now a function of $q_{\mathrm{max}}$ and we can see which value of $q_{\mathrm{max}}$ we need to satisfy the equation or how close or far we are from satisfying it. That $r_0$ is a measure of the range of the interaction in $r$-space is a well known feature of Quantum Mechanism (see Eq.~(2-42) of Ref.~\cite{pb:147pb}). 

Note that we have $\frac{\partial G}{\partial s}$ which we evaluate numerically. For this we evaluate Eq.~(\ref{2_3_1}) analytically, using the formula given in Ref.~\cite{Oller:1998hw}.

\subsection{Formalism for the nucleon nucleon interaction}
We follow the steps of the former subsection but we use the variable $E$ instead of $s$ ($E^2=s$). The $G$ function is now defined as 
\begin{align}\label{11_1}
   G(E)=\int_{|\textbf{q}|< q_{\mathrm{max}}}&\frac{d^3\textbf{q}}{(2\pi)^3}\frac{m_1m_2}{E_1(\textbf{q})E_2(\textbf{q})}\\\nonumber
   \times&\frac{1}{\sqrt{s}-E_1(\textbf{q})-E_2(\textbf{q})+i\epsilon}
\end{align}
It corresponds to Eq.~(\ref{2_3_1}) multiplied by $2m_12m_2$ for reasons of normalization of the fields and neglecting the negative energy parts of the relativistic propagator, as appropriate for heavy particles as the nucleons. In practice, we use the same formula of Ref.~\cite{Oller:1998hw} multiplied by $4m_1m_2$.

The potential is now
\begin{align}
   V=V_{0}+\beta(E-E_0) 
\end{align}
Then,
\begin{align}\label{T_baryoncheck}
    T=\frac{1}{[V_{0}+\beta(E-E_0)]^{-1} -G(E)}
\end{align}
The pole at $E_0$ implies
\begin{align}\label{11_1}
    V_{0}^{-1} -G(E_0)=0,\qquad V_{0}=\frac{1}{G(E_0)}
\end{align}
hence,
\begin{align}\label{11_2}
    T       =&\frac{1}{[\frac{1}{G(E_0)}+\beta(E-E_0)]^{-1} -G(E)}\\\nonumber
\end{align}
but now
\begin{align}
   \mathrm{Im}~G=-\frac{1}{8\pi}k\frac{2m_12m_2}{\sqrt{s}}
   =-\frac{1}{2\pi}k\frac{m_1m_2}{E}
\end{align}
and equivalently to Eq.~(\ref{8_4}) we have now
\begin{align}\label{11_3}
    &\frac{2\pi E}{m_1m_2}\bigg\{[\frac{1}{G(E_0)}+\beta(E-E_0)]^{-1} -\mathrm{Re}G(E)\bigg\}\\\nonumber
        \approx &\frac{1}{a}-\frac{1}{2}r_0k^2
\end{align}
Evaluating Eq.~(\ref{11_3}) at threshold we get 
\begin{align}\label{12_1}
    &\frac{2\pi E_{\mathrm{th}}}{m_1m_2}\bigg\{[\frac{1}{G(E_0)}+\beta(E-E_0)]^{-1} -\mathrm{Re}G(E_{\mathrm{th}})\bigg\}\\\nonumber
        =&\frac{1}{a}
\end{align} 
and using $E=E_1(k)+E_2(k)$, $E_i(k)=\sqrt{m_i^2+k^2}$,
\begin{align}
    \frac{\partial }{\partial k^2}&=\frac{\partial}{\partial E}\frac{\partial E}{\partial k^2}\\\nonumber
    &=(\frac{1}{2E_1(k)}+\frac{1}{2E_2(k)})\frac{\partial}{\partial E}\\\nonumber
    &=\frac{1}{2}\frac{E_1+E_2}{E_1(k)E_2(k)}\frac{\partial}{\partial E},
\end{align}
from the derivative of Eq.~(\ref{11_3}) at threshold we get
\begin{align}\label{12_2}
   \frac{2\pi}{m_1m_2}\bigg[&[\frac{1}{G(E_0)}+\beta(E_\mathrm{th}-E_0)]^{-1}\\\nonumber
    &-\mathrm{Re}G(E)_\mathrm{th}\bigg]\frac{E}{2E_1(k)E_2(k)}|_{E_{\mathrm{th}}}\\\nonumber
    +
    \frac{2\pi E_{\mathrm{th}}}{m_1m_2}\bigg[&-\beta[\frac{1}{G(E_0)}+\beta(E_\mathrm{th}-E_0)]^{-2}\\\nonumber
    &-\frac{\partial{\mathrm{Re}[G(E)]}}{\partial E}|_{E^{+}_{\mathrm{th}}}\bigg]\frac{E}{2E_1(k)E_2(k)}|_{E_{\mathrm{th}}}
    =-\frac{1}{2}r_0
\end{align}
or
\begin{align}\label{12_2}
   R_0\equiv -\frac{2\pi}{m_1m_2}\bigg[&[\frac{1}{G(E_0)}+\beta(E_\mathrm{th}-E_0)]^{-1} \\\nonumber
    &-\mathrm{Re}G(E_\mathrm{th})\bigg]\frac{m_1+m_2}{m_1m_2}\\\nonumber
    +
    \frac{2\pi (m_1+m_2}{m_1m_2})\bigg[&\beta[\frac{1}{G(E_0)}+\beta(E_\mathrm{th}-E_0)]^{-2}\\\nonumber
    &+\frac{\partial{\mathrm{Re}[G(E)]}}{\partial E}|_{E^{+}_{\mathrm{th}}}\bigg]\frac{m_1+m_2}{m_1m_2}
    =
    r_0
\end{align}
Once again we will compare $R_0$ versus $r_0$ as a function of $q_{\mathrm{max}}$. The couplings are now defined as  
\begin{align}
        g^2&=\lim\limits_{E\to E_0}(E-E_0)T\\\nonumber
    &=\lim\limits_{E\to E_0}\frac{E-E_0}{[\frac{1}{G(E_0)}+\beta(E-E_0)]^{-1} -G(E)}\\\nonumber
        &=\frac{1}{-[\frac{1}{G(E_0)}]^{-2}\beta-\frac{\partial G(E)}{\partial E}|_{E_0}}\\\nonumber
    &=\frac{1}{-G(E_0)^2\beta-\frac{\partial G(E)}{\partial E}|_{E_0}}
\end{align}
and 
\begin{align}\label{13_0}
  Z=1-X=-g^2G(E_0)^2\beta, ~X=-g^2\frac{\partial G(E)}{\partial E}|_{E_0}
\end{align}

The magnitude $\beta$ is now evaluated from Eq.~(\ref{12_1})

\begin{align}\label{13_1}
    &\beta=\\\nonumber
    &\frac{1}{E_{\mathrm{th}}-E_0}\bigg\{[\frac{1}{a}\frac{1}{2\pi}\frac{m_1m_2}{m_1+m_2}+\mathrm{Re}G(E_{\mathrm{th}})]^{-1}-\frac{1}{G(E_0)}\bigg\}
\end{align}
which determines $\beta$ from $a$, $E_0$ and $q_{\mathrm{max}}$. Substituted in Eq.~(\ref{12_2}), it allows to check $R_0$ versus $r_0$ as a function of $q_{\mathrm{max}}$.

\section{results}
\subsection{The deuteron case}
\begin{figure}[H]
  \centering
  \includegraphics[width=0.45\textwidth]{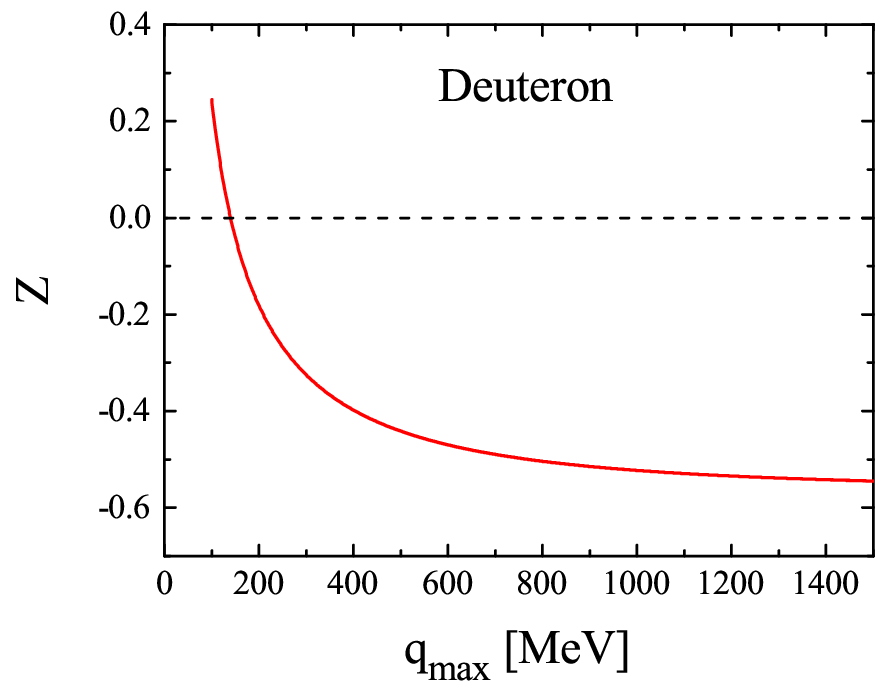}
  \caption{Z for the deuteron case as a function of $q_{\mathrm{max}}$.}\label{NNZ}
\end{figure}

We have the data of Eqs.~(\ref{2_3}). We determine $\beta$ in terms of this input and Eq.~(\ref{13_1}) and then the value of $Z$ of Eq.~(\ref{13_0}) as a function of $q_{\mathrm{max}}$ and plot $Z$ in Fig.~\ref{NNZ}

\begin{figure}[H]
  \centering
  \includegraphics[width=0.45\textwidth]{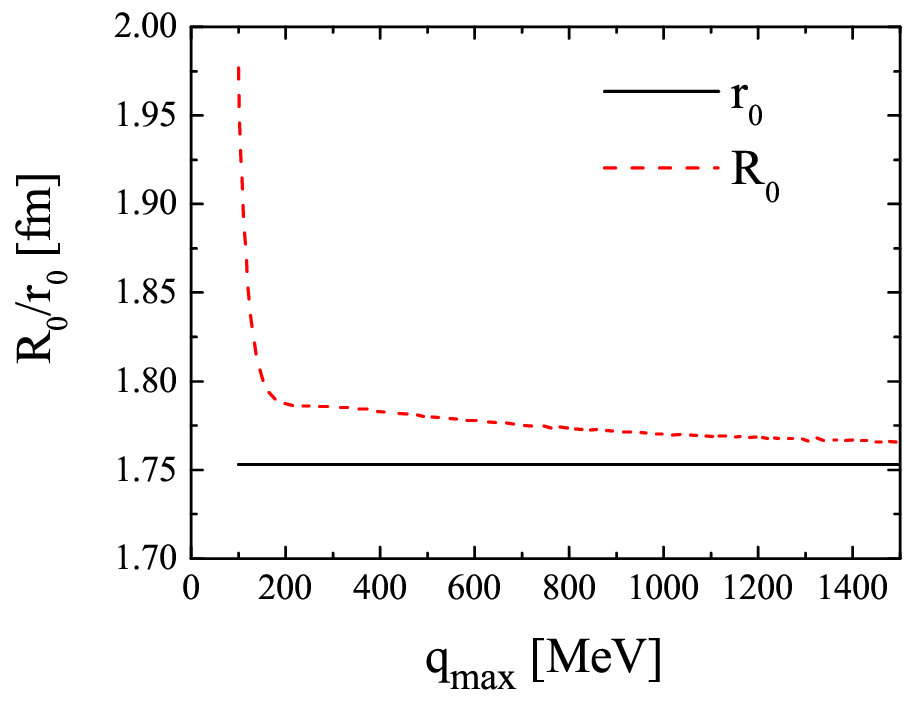}
  \caption{Comparison of $R_0$ and $r_0$ for the deuteron case as a function of $q_{\mathrm{max}}$.}\label{deotron_r}
\end{figure}

We can see in Fig.~\ref{NNZ}, that starting from $q_{\mathrm{max}}=100$~MeV, $Z$ is a small number, smaller than $0.25$ indicating a strong molecular $pn$ component. If we go beyond $q_{\mathrm{max}}=140$~MeV then $Z$ becomes negative and we should discard this situation. The comparison of the theoretical $R_0$ versus the experimental $r_0$ value is shown in Fig.~\ref{deotron_r}. There we can see that $R_0$ is close to $r_0$ for values of $q_{\mathrm{max}}\geq140$~MeV, but the disagreement becomes noticeable below this value. Altogether, there is a optimal situation around $q_{\mathrm{max}}\approx140$~MeV, where $Z\sim0$, indicating that the deuteron is a molecular state and the value of $R_0$ is very close to $r_0$. This situation is realistic indicating that the range of the $NN$ interaction in the deuteron in $r$-space is rather large, and hence $q_{\mathrm{max}}$ is small. If we look at the deuteron wave function in momentum space from the Bonn potential~\cite{Machleidt:2000ge}, we see that at $q=140$~MeV, the wave function is $7$\% the value at the origin and $\Phi^2(q=140~\mathrm{MeV})\approx0.006\Phi^2(0)$. We find then that the optimal agreement of the theory with the $a$ and $r_0$ data is obtained with a value of $q_{\mathrm{max}}$ which reflects realistically the actual deuteron wave function and tells us that $Z\sim0$, hence the deuteron is mostly a $pn$ molecule. There is no point demanding more than this information knowing that apart from the $S-$wave potential, there is also an important tensor part in the deuteron and a non negligible $D$-wave part in the wave function.
As we can see, the range of the interaction has been essential to obtain this acceptable picture. From Fig.~\ref{NNZ} we can see that if $q_{\mathrm{max}}$ increases, indicating short range interaction in $r$-space, $Z$ becomes negative and $X\approx1.6$ similar to what one obtains with the Weinberg formula of Eq.~(\ref{2_4}), not surprising since one implicitly is making  this assumption in the derivation of Weinberg formulas of Eqs.~(\ref{1_1}),~(\ref{1_2}) (see Ref.~\cite{Kinugawa:2021ykv}). Note that the consideration of the range of the interaction has also allowed us to obtain positive values of $R_0$ and close to the experimental one of $r_0$. Note also that in Eq.~(\ref{12_2}) the first term of $R_0$ is small and most of the contribution comes from the second term. If $\frac{\partial{\mathrm{Re}[G(E)]}}{\partial E}|_{E_{\mathrm{th}}}$ were zero, then $R_0$ would be negative and proportional to $Z$, as found in Eq.~(\ref{1_2}). Actually this is the situation with non relativistic kinematics when the range $q_{\mathrm{max}}$ is set to infinity (see non relativistic dimensional regularization results for $G$ in Ref~\cite{Kaplan:1996xu} where $\mathrm{Re}G=\mathrm{const}$ above threshold). In summary, the consideration of the range of the interaction has rendered us a picture of the deuteron far closer to the actual molecular nature than Weinberg's equations, and in return has shown that the interaction has to be of long range in $r$-space, with a realistic value of the range when compared to the actual deuteron wave function.
\section{The $D^*_{s0}(2317)$}
Next we pay attention to the $D^*_{s0}(2317)$ state considered as a $KD$ molecule. We will take the values of $a$, $r_0$ and the binding from the QCD lattice analysis of the finite volume levels of~\cite{MartinezTorres:2014kpc} (note $\frac{1}{a}$ in~\cite{MartinezTorres:2014kpc} versus $-\frac{1}{a}$ here in Eq.~(\ref{1_3}))
\begin{align}
    a(KD)&=+1.3\pm0.5\pm0.1~\mathrm{fm}\\\nonumber
    r_0(KD)&=-0.1\pm0.3\pm0.1~\mathrm{fm}    
\end{align}
and we take the nominal mass $2317$~MeV for the mass of the state.

\begin{figure}
  \centering
  \includegraphics[width=0.45\textwidth]{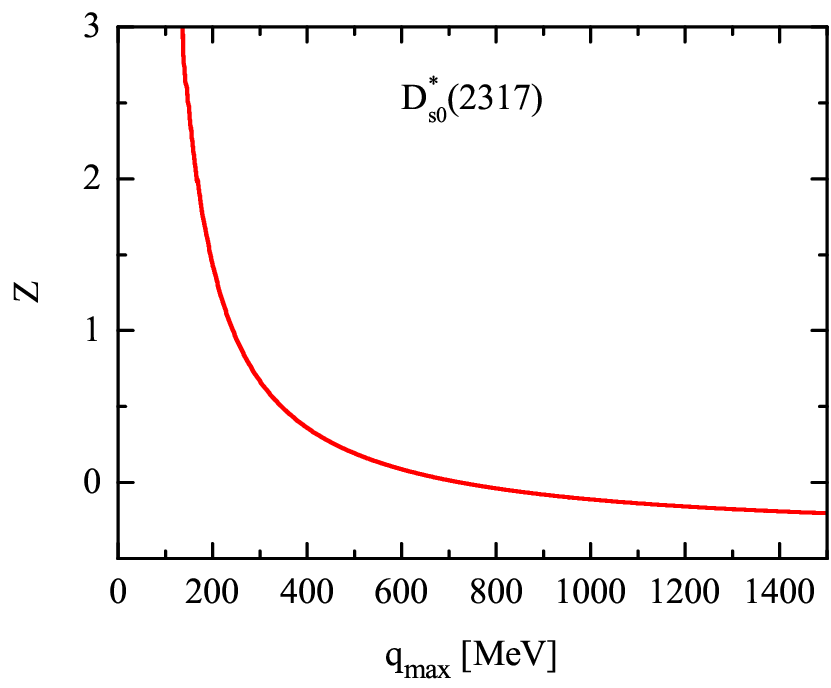}
  \caption{The value of $Z$ for $KD$ scattering forming the $D^{*}_{s0}(2317)$.}\label{Z_2317}
\end{figure}

\begin{figure}
  \centering
  \includegraphics[width=0.45\textwidth]{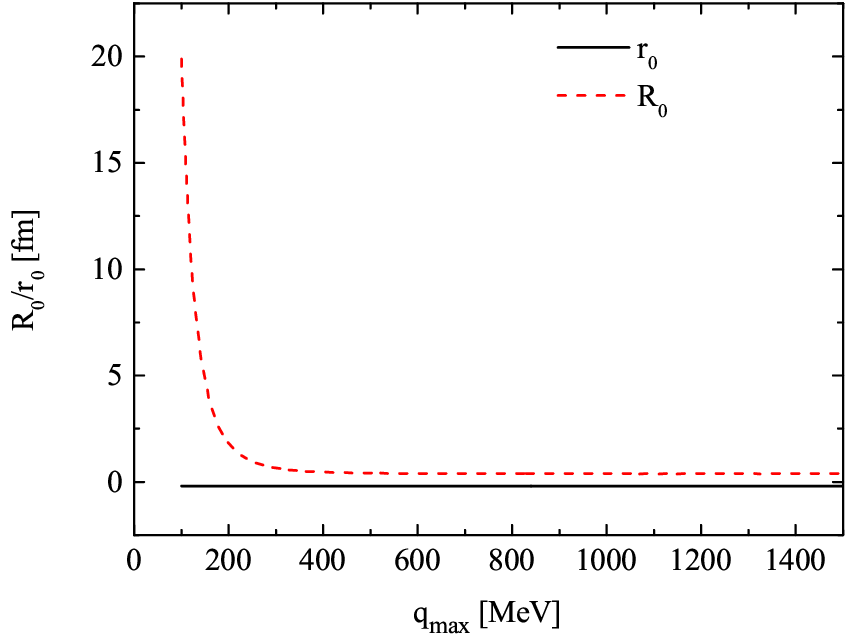}
  \caption{The comparison of $R_0$ and $r_0$ for the $D^{*}_{s0}(2317)$ case as a function of $q_{\mathrm{max}}$.}\label{2317_r}
\end{figure}

In Fig.~\ref{Z_2317} we see that $Z$ takes unrealistically large values for $q_{\mathrm{max}}< 300$~MeV. On the other hand, in Fig.~\ref{2317_r} we plot $R_0$ versus $r_0$ and we see that for $q_{\mathrm{max}}< 300$~MeV, the deviation of $R_0$ and $r_0$ becomes gradually large and unacceptable. We could say that for values of $q_{\mathrm{max}}> 400$~MeV we already obtain an acceptable agreement of $R_0$ versus $r_0$. We see then from Fig.~\ref{Z_2317} that in this case $Z< 0.4$, indicating a $DK$ molecular component with probability larger than $60$\%. This would be in agreement with the findings in~\cite{MartinezTorres:2014kpc} where it was found that 
\begin{align}
    P(DK)=(72\pm13\pm5)\%
\end{align}

We also see that the range, with $q_{\mathrm{max}}\ge 400$~MeV, corresponds to a shorter range in $r$-space than in the deuteron case. In the dynamical picture of~\cite{Gamermann:2006nm} the interaction is driven by vector meson exchange, contrary to the $np$ interaction where pion exchange plays a dominant role. 

Note also from Fig.~\ref{Z_2317} that around $q_{\mathrm{max}}= 725$~MeV $Z$ becomes zero and negative from there on. This also gives us an idea of the range demanded by the data, which reflects realistically what one can expect from light vector exchange. From comparison recall that $q_{\mathrm{max}}=630$~MeV was used in the study of the $\Bar{K}N$ interaction in~\cite{Oset:1997it}.

\section{The $D^*_{s1}(2460)$}
\begin{figure}
  \centering
  \includegraphics[width=0.45\textwidth]{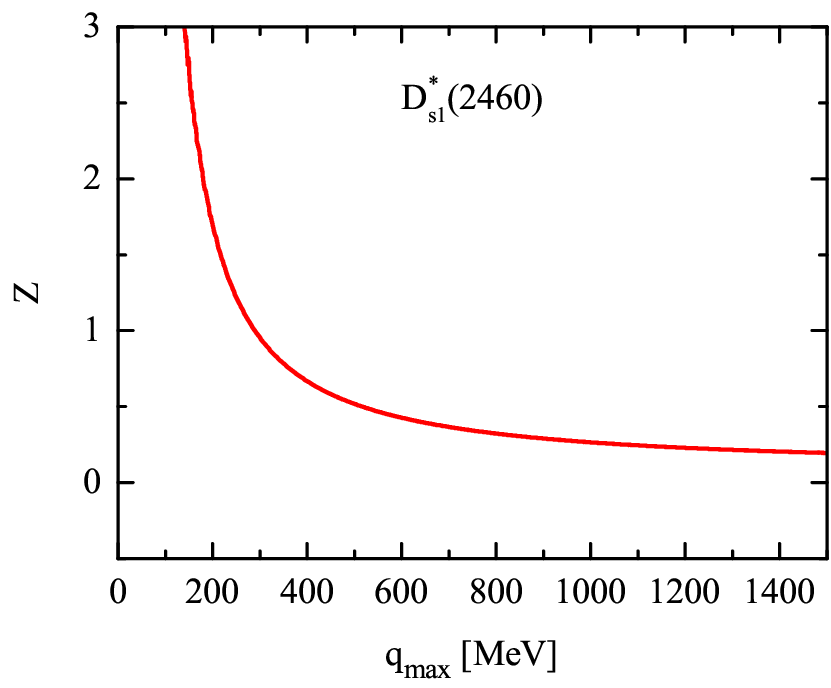}
  \caption{Z for the $D^{*}_{s1}(2460)$ case as a function of $q_{\mathrm{max}}$.}\label{Z_2460}
\end{figure}

\begin{figure}
  \centering
  \includegraphics[width=0.45\textwidth]{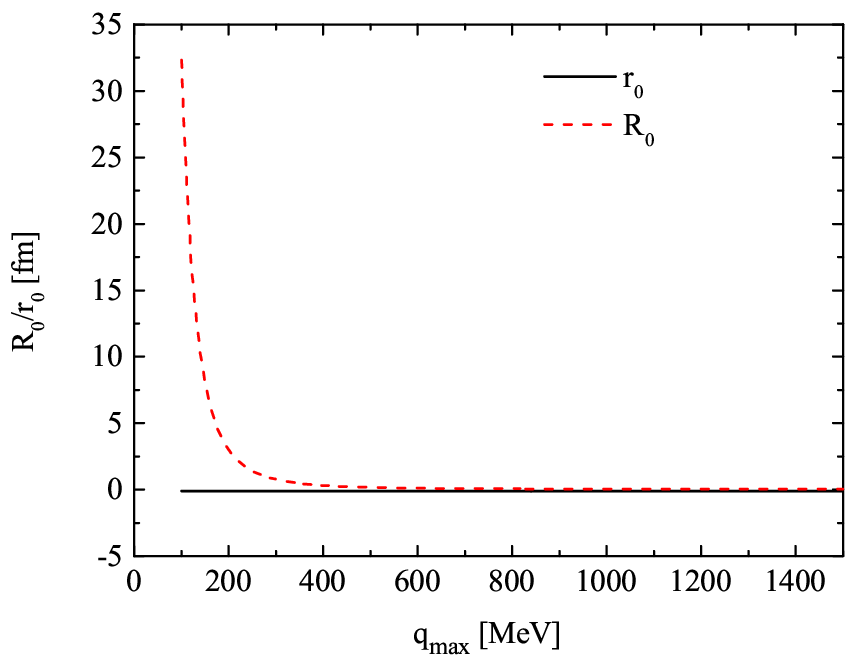}
  \caption{The comparison of $R_0$ and $r_0$ for the $D^{*}_{s1}(2460)$ case as a function of $q_{\mathrm{max}}$.}\label{2460_r}
\end{figure}

Now we analyse the $D^*_{s1}(2460)$ state shown to be mostly molecular in the $KD^{*}$ channel from the analysis of the lattice QCD  levels in~\cite{MartinezTorres:2014kpc}. There it was found
\begin{align}
    a(KD^*)&=+1.1\pm0.5\pm0.2~\mathrm{fm}\\\nonumber
    r_0(KD^*)&=-0.2\pm0.3\pm0.1~\mathrm{fm}    
\end{align} 
and we take the nominal mass $2460$~MeV for the state. In Fig.~\ref{Z_2460} we show the value of $Z$ as a function of $q_{\mathrm{max}}$. We have to go to values of $q_{\mathrm{max}}$ bigger than $280$~MeV to have $Z$ smaller than $1$. We can complement this information from Fig.~\ref{2460_r} which shows a big discrepancy of $R_0$ with $r_0$ for values of $q_{\mathrm{max}}$ smaller than $400$~MeV.

We can see now that $Z$ never becomes zero, independent of $q_{\mathrm{max}}$, reaching a value of $0.2$ for large $q_{\mathrm{max}}$. If we take a range of $q_{\mathrm{max}}$ like in the former case $400~\mathrm{MeV}<q_{\mathrm{max}}<750~\mathrm{MeV}$, the $Z$ range becomes $0.3< Z< 0.6$. What we can see is that $Z$ is now bigger than in the case of the $D^*_{s0}(2317)$. The compositeness is now around or bigger than $40$\% ($0.4< X< 0.7$) in agreement with the findings of~\cite{MartinezTorres:2014kpc}, where one obtains 
\begin{align}
    P(KD^*)=(57\pm21\pm6)\%
\end{align}
In this case the $\eta D^*_{s}$ channel is mostly responsible for the remaining probability. It is interesting to see that the analysis done here renders values of $Z$ which are in good agreement with those found in~\cite{MartinezTorres:2014kpc}.

What we have found in the three cases studied is that the range of the interaction is very important to consider, even in cases little bound like the deuteron. We could see that the combined information of $a$, $r_0$ and the binding could provide a fair information on the $D^*_{s0}(2317)$ and $D^*_{s1}(2460)$ which are bound by about $40-45$~MeV. At the same time the analysis gives us some idea about the range of the interaction, with the $NN$ interaction for the deuteron being of longer range in $r$-space than the $KD$ and $KD^*$ in the cases of the $D^*_{s0}(2317)$ and $D^*_{s1}(2460)$ states. The information obtained, even with its uncertainties, is more accurate than that obtained by the limiting equations of (\ref{1_1}), (\ref{1_2}). The algorithms resulting from our study are also easy to implement and can be used to study other cases. 

\section{conclusion}
We have used a formalism for the scattering of hadrons using a separable potential showing explicitly the range of the interaction. We showed the equivalence of the formalism to the one used in the chiral unitary approach with the on shell factorization of the potential. The success of this latter approach generating low lying resonances and describing many physical process gives us confidence in the method used in the present work. When studying the scattering of the two hadrons sometimes a bound state appears below threshold and it is logical to ask oneself whether that state corresponds to a molecular state of these hadronic components, stemming from this interaction, or it corresponds to a different structure. Sometimes the state could have a mixture of another pair of hadrons, or even have a component of a compact quark cluster. To account for all these cases we have assumed an energy dependent potential, and our formalism allows one to determine the scattering length and the variable $q_{\mathrm{max}}$, the range of the interaction in momentum space. The formalism also allows one to obtain the effective range $r_0$, and not always is it possible to get agreement with experiment, indicating that more information beyond the effective range expansion would be necessary for a better analysis of the data. Yet, the comparison of the theoretical value $R_0$ and $r_0$  shows that there are regions of  $q_{\mathrm{max}}$ where the disagreement is too big and unacceptable. Looking at the regions of  $q_{\mathrm{max}}$ where $R_0$ is closer to $r_0$ one can find a double information, which is the range of the interaction and the value of the $Z$, or the molecular compositeness $X=1-Z$. The combined analysis using the information of the binding, $a$ and $r_0$ renders us reasonable values of $Z$ and the range of the interaction for the three systems studied, with very different binding energy: the deuteron with $2.22$~MeV binding and the $D^*_{s0}(2317)$ and $D^*_{s1}(2460)$ states with binding around $40-45$~MeV. The information obtained with this method is more accurate than that obtained from the formalism of Weinberg, derived in the limit of very small binding and zero range of the interaction in $r$-space.

At this point it is worth going back to the assumptions made and how one could improve on what has been done here. First, let us state clearly that we do not get a precise value of $Z$ and the range of the interaction from the values of $a$ and $r_0$. We obtain qualitatively a band of values for the $Z$ and $q_{\mathrm{max}}$, yet a very valuable information. If one wished to go further on what we have done, one could try to see if different result come with another formalism which does not involve a separable potential, although, for the reasons discussed, the approach looks very reliable. Eventually one could try to have a potential which is not linear in $s$ as we have assumed here, for which we provided information suggesting that it is quite a good assumption. Obviously, following these steps one would introduce more free parameters. The idea here is to see how much one can learn from $a$ and $r_0$ alone and we had two free parameters, $q_{\mathrm{max}}$ and $\beta$ to match to $a$ and $r_0$. Our believe is, indeed, that if one wishes to learn more about $Z$ and  $q_{\mathrm{max}}$, one would have to use more data on scattering, or other processes, that allow one to go beyond the effective range expansion. This is certainly a commendable task. Yet, the point here was to see how much one can learn from $a$ and $r_0$ alone, and we showed that one can get some qualitative knowledge about the values of $Z$ that are more accurate than the values provided by the standard Weinberg formalism, and at the same time one gets an additional information on the range of the interaction.

\section{Acknowledgements}
We would like to thank Juan Nieves and Miguel Albaladejo for useful discussions. This work is partly supported by the National Natural Science Foundation of China under Grants Nos. 11975009, 12175066 and 12147219. This work is also partly supported by the Spanish Ministerio de Economia y Competitividad (MINECO) and European FEDER funds under Contracts No. FIS2017-84038-C2-1-PB, PID2020-112777GB-I00, and by Generalitat Valenciana under contract PROMETEO/2020/023. This project has received funding from the European Union Horizon 2020 research and innovation programme under the program H2020-INFRAIA-2018-1, grant agreement No. 824093 of the STRONG-2020 project. One of us, Jing Song wishes to acknowledge support from China Scholarship Council.

\bibliography{refs.bib}
\end{document}